\documentclass{iaus}
\usepackage{graphicx}
\usepackage{epsfig}
\usepackage{amssymb}
\usepackage{natbib}

\title[Stellar Ages from Stellar Rotation] 
{Stellar Ages from Stellar Rotation}

\author[S{\o}ren Meibom]   
{S{\o}ren Meibom}

\affiliation{Harvard-Smithsonian Center for Astrophysics\\
60 Garden Street, Cambridge, MA, 02138, USA\\
email: {\tt smeibom@cfa.harvard.edu}}

\pubyear{2009}
\volume{xxx}  
\pagerange{XXX--YYY}
\jname{The Ages of Stars}
\editors{A.C. Editor, B.D. Editor \& C.E. Editor, eds.}
\begin{document}

\maketitle


\begin{abstract}
Our ability to determine stellar ages from measurements of stellar
rotation, hinges on how well we can measure the dependence of rotation
on age for stars of different masses. Rotation periods for stars in
open clusters are essential to determine the relations between stellar
age, rotation, and mass (color). Until recently, ambiguities in $v\sin i$
data and lack of cluster membership information, prevented a clear empirical
definition of the dependence of rotation on color. Direct measurements
of stellar rotation periods for members in young clusters have now revealed
a well-defined period-color relation. We show new results for the open
clusters M35 and M34.
However, rotation periods based on ground-based observations are limited
to young clusters. The Hyades represent the oldest coeval
population of stars with measured rotation periods. Measurements of
rotation periods for older stars are needed to properly constrain the
dependence of stellar rotation on age. We present our plans to use
the Kepler space telescope to measure rotation periods in clusters
as old as and older than the Sun.
\end{abstract}


\section{Introduction}

Knowing stellar ages is fundamental to understanding the time-evolution
of various astronomical phenomena related to stars and their companions.
Accordingly, over the past decades much work has been focused on identifying
the properties of a star that best reveal its age. For coeval populations
of stars in clusters, the most reliable ages are determined by fitting model
isochrones to single cluster members in the color-magnitude
diagram. However, for the vast majority of stars not in clusters
(unevolved late-type field stars), ages determined using the isochrone method
are highly uncertain because the primary age indicators are nearly constant
throughout their main-sequence lifetimes, and because their distances and
thus luminosities are poorly known. Therefore, finding a distance-independent
property of individual stars that can act as a reliable determinant of their
ages will be of great value.

Stellar rotation (and the related measure of chromospheric activity - see
paper by Mamajek this volume) has emerged as a promising and distance-
independent indicator of age
\citep[e.g.][]{skumanich72,kawaler89,barnes03a,barnes07}.
\citet{skumanich72} first established stellar rotation as an astronomical
clock by relating the average projected rotation velocity in
young open clusters to their ages via the expression $\overline{v\sin i}
\propto t^{-0.5}$. The Skumanich relation is limited in mass to
early G dwarfs and suffers from the ambiguity (due to the unknown inclination
angle) of the $v\sin i$ data. Furthermore, for ages beyond that of the
Hyades cluster ($\sim625$Myr), the Skumanich relationship is constrained
only by a single G2 dwarf - the Sun.

Modern photometric time-series surveys in young open clusters can provide
precisely measured stellar rotation periods (free of the $\sin i$ ambiguity)
for F, G, K, and M dwarfs. Based on such new data and emerging empirical
relationships between stellar rotation, color, and age, a method was proposed
by \citet{barnes03a} to derive ages for late-type dwarfs from observations
of their colors and rotation periods alone. We refer the reader to the paper
in this volume by Barnes for a motivation and description of the method of
{\it gyrochronology}. However, our ability to determine stellar ages from
stellar rotation, hinges on how well we can measure the dependence of 
rotation on age for stars of different masses.


\section{The key role of open clusters}

As coeval populations of stars with a range of masses and well determined
ages, open cluster fulfill a critically important role in calibrating the
relations between stellar age, rotation, and color. Indeed, {\it open
clusters can define a surface in the 3-dimensional space of stellar
rotation period, color, and age, from which the latter can be determined
from measurements of the former two} (see Figure~\ref{3d_car} below).

This inherent quality of open clusters can only be fully exploited if
precise stellar rotation periods (free of the $\sin i$ ambiguity) are
measured for cluster members.  Accordingly, the time base-line and
frequency of time-series photometric observations must be long enough
and high enough, respectively, to avoid a bias against detecting periods
of more slowly rotating stars, and to avoid detection of false rotation
periods due to aliases and a strong ``window-function'' in the data.
Furthermore, measured rotation periods should be combined with
information about cluster membership and multiplicity. Removing
non-members and stars in close binaries affected by tidal
synchronization will allow a better definition of the relationship
between rotation period and color at the age of the cluster.
Finally, identification of single cluster members will enable a better
cluster age to be determined from isochrone fitting. The new results
for the open clusters M35 and M34
shown in Figure~\ref{m3534pbv}, reflect the powerful combination of
decade-long time-series spectroscopy for cluster membership and
time-series photometry over 5 months for stellar rotation periods.


\section{New observations in the open clusters M35 and M34}

We carried out photometric monitoring campaigns over 5 consecutive months
for rotational periods, and nearly decade-long radial-velocity surveys for
cluster membership and binarity, on the $\sim$150\,Myr and $\sim$200\,Myr
open clusters M35 and M34. For detailed descriptions of the observations,
data-reduction, and data-analysis, see \citet{mm05,mms06,mms08},
and \citet{bmm09}.

\begin{figure}[ht!]
\includegraphics[height=.33\textheight]{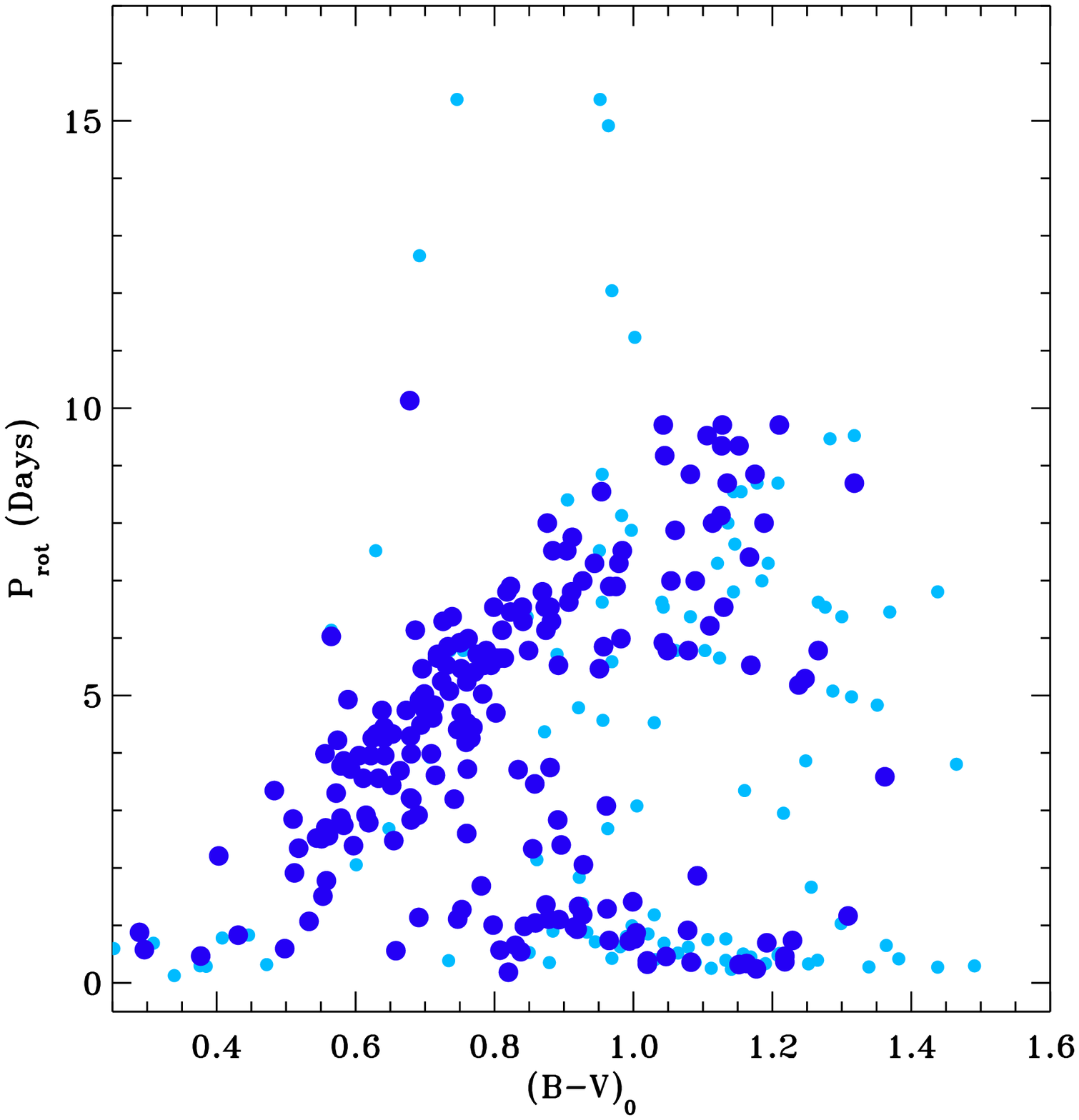}
\includegraphics[height=.33\textheight]{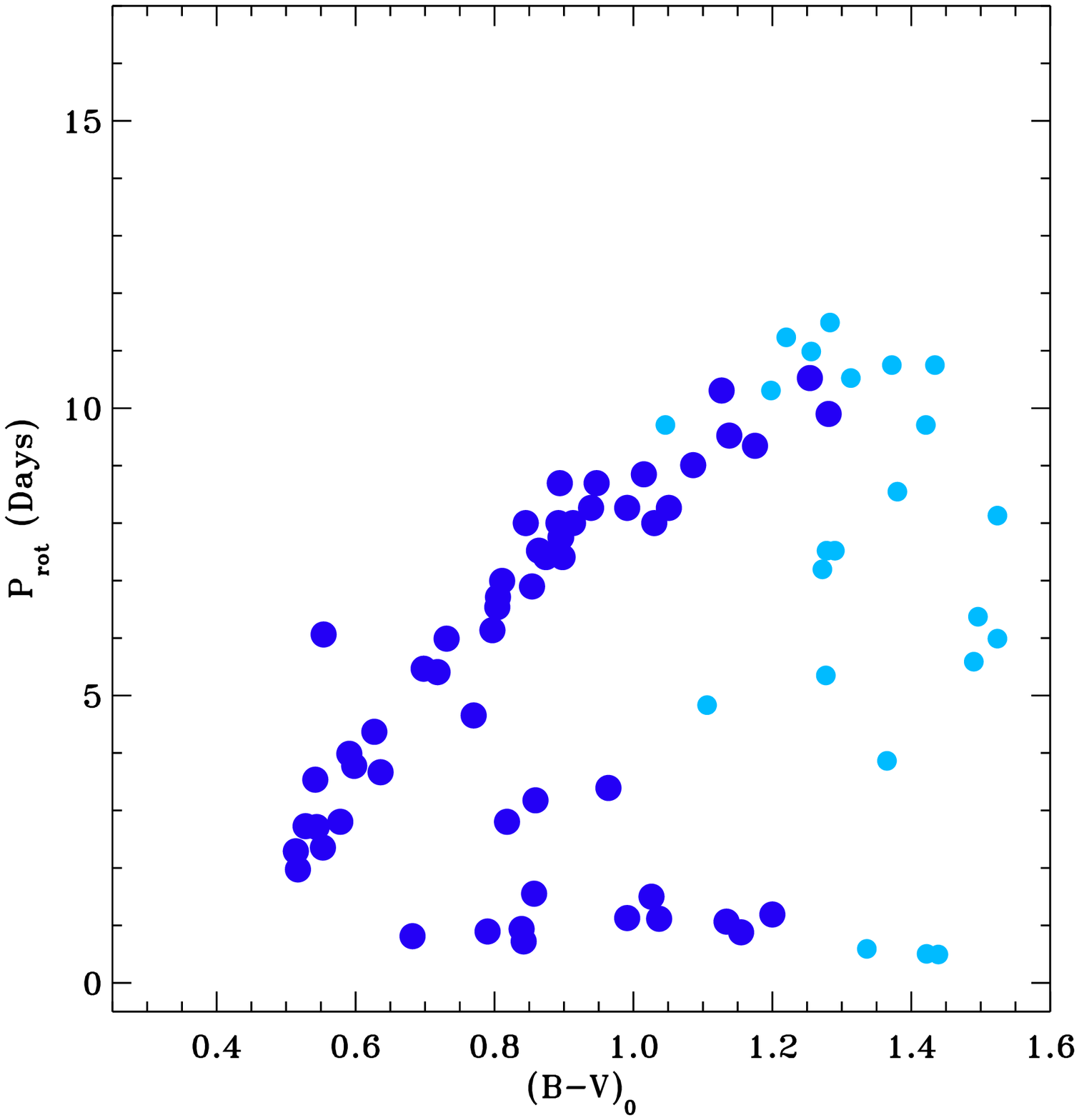}
\caption{The distribution of stellar rotation periods with (B-V)
color index for 310 members of M35 ({\it left}; \citep{mms08}) and
79 members of M34 ({\it right}). Dark blue (black) plotting symbols
are used for radial-velocity members and light blue (grey) for
photometric members.}
\label{m3534pbv}
\end{figure}

{\it Time-Series Photometric Observations}:
We surveyed, over a timespan of 143 days, a region of $40 \times 40$ arc
minutes centered on each cluster. Images were acquired at a frequency of
once a night both before and after a central block of 16 full nights with
observations at a frequency of once per hour. The data were obtained in
the Johnson V band with the WIYN 0.9m telescope on Kitt Peak. Instrumental
magnitudes were determined from Point Spread Function photometry. Light
curves were produced for more than 14,000 stars with $12 < V < 19.5$.
Rotational periods were determined for 441 and 120 stars in the fields
of M35 and M34, respectively (see Figure~\ref{m3534pbv}).

{\it The spectroscopic surveys}:
M35 and M34 have been included in the WIYN Open Cluster Study
(WOCS; \citet{mathieu00}) since 1997 and 2001. As part of WOCS, 1-3
radial-velocity measurements per year were obtained on both clusters
within the 1-degree field of the WIYN 3.5m telescope with the multi-object
fiber positioner (Hydra) feeding a bench-mounted echelle spectrograph.
Observations were done at central wavelengths of 5130\AA\ or 6385\AA\
with a wavelength range of $\sim$200\AA\ . From this spectral region
with many narrow absorption lines, radial velocities were determined
with a precision of $< 0.4~$km/s \citep{gmh+08,mbd+01}. Of the stars
with measured rotational periods in M35 and M34, 203 and 56,
respectively, are radial-velocity members of the clusters (dark blue
symbols in Figure~\ref{m3534pbv}). Including photometric members
(light blue symbols in Figure~\ref{m3534pbv}), the total number of
stars with measured rotational periods in M35 and M34, are 310
and 79.

\section{The color-period diagram}

Figure~\ref{m3534pbv} shows the rotational periods for members in
M35 and M34 plotted against their dereddened $B-V$ colors. The coeval stars
fall along two well-defined sequences representing two different
rotational states. One sequence displays a clear correlation between
rotation period and color, and forms a diagonal band of stars whose periods
are increasing with increasing color index (decreasing mass). The
second sequence consists of rapidly rotating stars and shows little
mass dependence. A small subset of stars is distributed
between the two sequences. The distribution of stars in the color-period
diagrams suggests that the rotational evolution is slow where we
see the sequences and fast in the gap between them. Other
areas of the color-period plane are either unlikely or ``forbidden''.


\section{The dependence of stellar rotation period on color}

For our purpose of determining the dependence of stellar rotation on
stellar color, we can focus on the diagonal sequence of more slowly
rotating stars in Figure~\ref{m3534pbv}. We can do so because surveys
for stellar rotation in the older clusters M37 (550\,Myr) and the
Hyades (625\,Myr) show that F, G, and K dwarfs spin down over a few
hundred million years and converge onto this sequence \citep{hgp+08,rtl+87}.

Barnes (2003, 2007, and this volume) refer to the diagonal sequence
as the Interface (I) sequence and propose a function
($f(B-V)$) to represent it \citep{barnes07}:

\begin{equation}
P(t,B-V) = g(t) \times f(B-V)
\end{equation}

\noindent where 

\begin{equation}
f(B-V) = a((B-V)-b)^{c}
\end{equation}

~\\
\noindent with $a = 0.77$ and $c = 0.60$. \citet{barnes07} fix $b$
at a value of 0.4, and determine $g(t) = t^{0.52}$.

From the method of gyrochronology \citep{barnes03a,barnes07}, the
functional dependence between stellar color and rotation period
($f(B-V)$) will directly affect the derived ages, and will, if not
accurately determined, introduce a systematic error. It is therefore
important to constrain and test the color-rotation relation for stars
on the I sequence as new data of sufficiently high quality becomes
available. \citet{mms08} fit $f(B-V)$ as given in equation [5.2] to
the I sequence stars in M35, leaving all 3 coefficients ($a$, $b$, $c$)
as free parameters.
They get the same value of 0.77 for $a$, but a slightly different value
of 0.55 for $c$. By leaving the translational term $b$ free, a value
of 0.47 was found. This value for $b$ is interesting because it corresponds
to the approximate $B-V$ color for F-type stars at the transition from
a radiative to a convective envelope. This transition is also associated
with the onset of
effective magnetic wind breaking \citep[e.g.][]{schatzman62}, and known
as the break in the Kraft curve \citep{kraft67}. The value of 0.47 for
the $b$ coefficient therefore suggest that, for M35, the blue (high-mass)
end of the I sequence begins at the break in the Kraft curve.

The I sequence in M34 is particularly well-defined and will be used
to further constrain the dependence between rotation and color in a
forthcoming paper (Meibom et al. 2009, in preparation).


\section{The dependence of stellar rotation on age}

With well-defined color-rotation relations (I sequences) for clusters
of different ages, we are able to constrain the dependence of stellar
rotation on age for stars of different masses. 
Comparison of the rotation periods for F-, G-, and K-type I sequence
dwarfs at different ages enable a direct test of the Skumanich
relationship for early G dwarfs and for dwarfs of higher and lower masses.

Initial comparisons in \citet{mms08} between the rotation periods of
G and K dwarfs on the I sequences in M35 and the Hyades, suggest that
the Skumanich time-dependence ($P_{rot} \propto t^{0.5}$) can account
for the evolution in rotation periods between M35 and the Hyades for
G dwarfs. However, the time-dependence for spin-down of K dwarfs is
different and slower than Skumanich.
In a more in depth analysis (preliminary results) Meibom et al. (2009;
in preparation) calculate the mean rotation periods for late-F, G, early
K, and late-K I sequence dwarfs in M35, M34, NGC3532 (Barnes 2003; 300\,Myr),
M37, and the Hyades. They find that the increase in the mean rotation period
with age is consistent with Skumanich spin-down for the late-F and G dwarfs,
whereas K dwarfs spin down significantly slower.
This deviation from Skumanich spin-down for
K dwarfs, suggest that the rotation period for late-type stars cannot be
expressed as the product of separable functions of time and color
(Eq. [5.1]).  Skumanich spin-down was assumed for all (late-F through
early M) stars in \citet{barnes03a,barnes07} and in \citet{kawaler89}.

Eventually, when rotation periods of sufficient quality is available
for a larger number of clusters, the effects on the rotational evolution
of other stellar parameters, e.g. metallicity, and of the cluster
environment, should be considered. Irwin, this volume, give a more complete
list of published rotation data in clusters.


\section{The Kepler mission - a unique opportunity}

\begin{figure}[ht!]
\begin{center}
\includegraphics[width=5.5in]{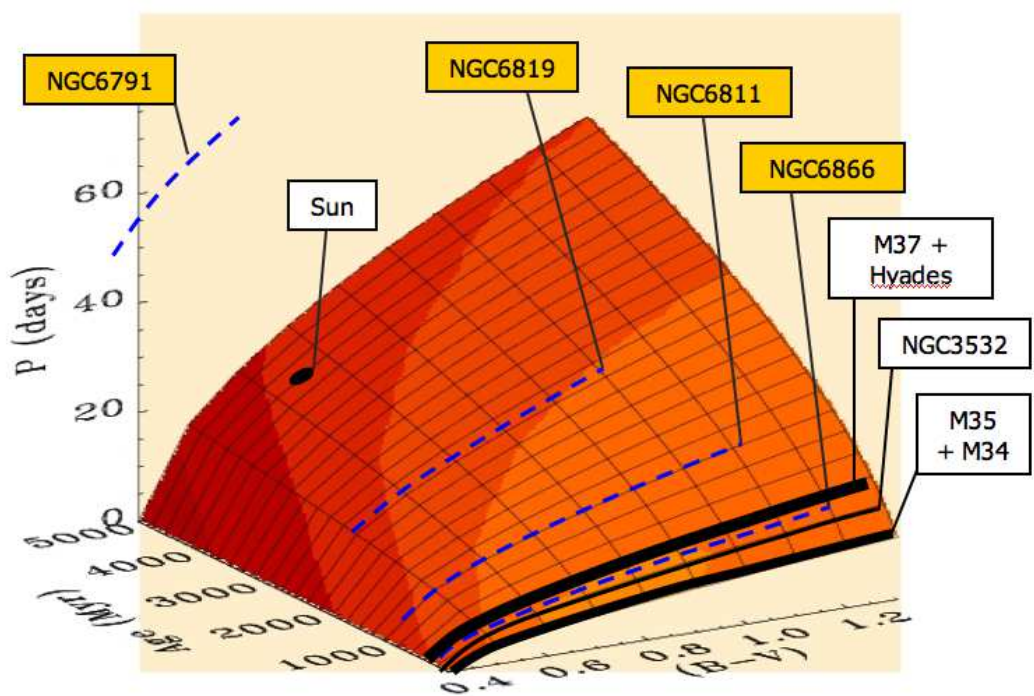} 
\caption{
A schematic of the (presumed) empirical surface in the 3-dimensional
parameter space of stellar age (Myr), color, and rotation period.
The surface is currently defined {\it only} by stars in young open
clusters (black solid lines), and by the Sun (black dot). The dashed
blue lines mark the ages and approximate color ranges of FGK dwarfs
in the 4 open clusters within the Kepler field.}
\label{3d_car}
\end{center}
\end{figure}

At the present time, the Hyades represent the oldest coeval population
of stars with measured rotation periods. Measurements of rotation periods
for older late-type dwarfs is needed to properly constrain the dependence
of stellar rotation on age and mass, and to calibrate the technique of
gyrochronology. Figure~\ref{3d_car} shows a schematic of the surface
in the 3-dimensional space of rotation, color, and age. At the present
time this surface is defined solely by color-period data in young
clusters and for the Sun.
The solid black curves represent the ages and color-ranges of FGK
dwarfs in M35, M34, NGC3532, M37, and the Hyades. The color
and age of the Sun is marked as a solid dot. The figure demonstrates
clearly the need for observations of stellar rotation periods beyond
the age of the Hyades.

The lack of periods for older stars (with the exception of the Sun)
reflects the challenging task of measuring - from the ground - photometric
variability for slowly rotating stars with ages of $\sim$1Gyr or more.
However, the Kepler space telescope (scheduled for a 2009 launch),
will provide photometric measurements with a precision,
cadence, and duration, sufficient to
measure stellar rotation periods from brightness modulations for stars
as old as and older than the Sun. Four open clusters are located within
the Kepler target region: NGC\,6866 ($\sim$0.5\,Gyr), NGC\,6811
($\sim$1\,Gyr), NGC\,6819 ($\sim$2.5\,Gyr), and NGC\,6791 ($\sim$10\,Gyr).
With Kepler we therefore have a unique opportunity to extend the
age-rotation-color relationships beyond the age of the Hyades and
the Sun. The dashed blue curves in Figure~\ref{3d_car} mark the
ages and approximate color ranges of FGK dwarfs in the 4 clusters. 



\def\aj{AJ}\def\araa{ARA\&A}\def\apj{ApJ}\def\apjl{ApJ}\def\apjs{ApJS}\def\ao{%
Appl.~Opt.}\def\apss{Ap\&SS}\def\aap{A\&A}\def\aapr{A\&A~Rev.}\def\aaps{A\&AS}%
\def\azh{AZh}\def\baas{BAAS}\def\jrasc{JRASC}\def\memras{MmRAS}\def\mnras{MNRA%
S}\def\pra{Phys.~Rev.~A}\def\prb{Phys.~Rev.~B}\def\prc{Phys.~Rev.~C}\def\prd{P%
hys.~Rev.~D}\def\pre{Phys.~Rev.~E}\def\prl{Phys.~Rev.
  Lett.}\def\pasp{PASP}\def\pasj{PASJ}\def\qjras{QJRAS}\def\skytel{S\&T}\def\s%
olphys{Sol.~Phys.}\def\sovast{Soviet~Ast.}\def\ssr{Space~Sci.~Rev.}\def\zap{ZA%
p}\def\nat{Nature}\def\iaucirc{IAU~Circ.}\def\aplett{Astrophys.~Lett.}\def\aps%
pr{Astrophys.~Space~Phys.~Res.}\def\bain{Bull.~Astron.~Inst.~Netherlands}\def\%
fcp{Fund.~Cosmic~Phys.}\def\gca{Geochim.~Cosmochim.~Acta}\def\grl{Geophys.~Res%
.~Lett.}\def\jcp{J.~Chem.~Phys.}\def\jgr{J.~Geophys.~Res.}\def\jqsrt{J.~Quant.%
~Spec.~Radiat.~Transf.}\def\memsai{Mem.~Soc.~Astron.~Italiana}\def\nphysa{Nucl%
.~Phys.~A}\def\physrep{Phys.~Rep.}\def\physscr{Phys.~Scr}\def\planss{Planet.~S%
pace~Sci.}\def\procspie{Proc.~SPIE}

\end{document}